\documentclass[a4paper,11pt]{article}

\usepackage[margin=25mm]{geometry}
\usepackage{amsmath,amssymb}
\usepackage{booktabs}
\usepackage{array}
\usepackage{longtable}
\usepackage{graphicx}
\usepackage{float}
\usepackage{setspace}
\usepackage{enumitem}
\usepackage{hyperref}
\usepackage{xurl}

\hypersetup{
    colorlinks=false,
    pdfborder={0 0 0}
}

\hypersetup{
    colorlinks=false,
    pdfborder={0 0 0}
}

\begin{document}

\vspace*{-10mm}

\begin{center}

{\large\bfseries
Price Responses of Rwandan Tungsten Exports under Conflict Minerals Regulation\\
---Period-by-Period Estimation of Export Demand Elasticity---
}

\vspace{4mm}

{\normalsize
Haruka Nagamori$^{1}$,  Kazuhiko Nishimura$^{2}$ \\ Graduate School of Humanities and Social Sciences, Chukyo University$^{1}$ \\
Institute of Economics, Chukyo University$^{2}$

E-mail: u62501m@m.chukyo-u.ac.jp$^{1}$, nishimura@lets.chukyo-u.ac.jp$^{2}$
}

\vspace{2mm}

{\normalsize
August 8, 2026
}

\end{center}

\vspace{4mm}

% Hide page numbers
\pagestyle{empty}

% Paragraph settings
\setlength{\parindent}{1em}
\setlength{\parskip}{0pt}

%==============================
% Main text
%==============================

Section 1502 of the Dodd--Frank Act, enacted in 2010, requires U.S.-listed companies that use tin, tantalum, tungsten, and gold (3TG) originating from the Democratic Republic of the Congo and adjoining countries to disclose information such as the minerals' countries of origin. At the same time, it has been pointed out that avoidance of sourcing from the covered region may have resulted in a de facto embargo. However, it remains insufficiently understood how the price responsiveness of mineral exports evolved before and after the implementation of the regulation and under subsequent changes in the institutional and market environment. Among the 3TG minerals, tungsten production in the covered region is almost entirely concentrated in Rwanda. This study therefore focuses on Rwandan tungsten and estimates export demand price elasticities by period from January 2009 to December 2023.

Because missing export quantity data prevent export unit values from being observed continuously, this study applies the identification approach of Nakano and Nishimura (2025). Monthly mirror trade data from UN Comtrade are combined with exchange rates and a world average price, and an importer fixed-effects model is estimated with export value as the dependent variable. The sample period is divided into four periods based on changes in the institutional and market environment.

The results show a statistically significant negative price response in Period 1 ($\eta=-20.814$, $p<0.01$). No clear price response is observed in Period 2 ($\eta=1.814$, $p>0.10$). A statistically significant negative response reappears in Period 3 ($\eta=-5.277$, $p<0.01$), whereas no clear response is observed in Period 4 ($\eta=0.440$, $p>0.10$). In addition, coefficient-difference tests identify statistically significant differences between Period 1 and Period 2 ($p=0.0021$) and between Period 3 and Period 4 ($p=0.0006$). These findings indicate that the price responsiveness of Rwandan tungsten exports did not change in only one direction after the introduction of the regulation, but continued to vary over time.

\vspace{4mm}

%------------------------------
% Keywords
% Up to five terms
%------------------------------
{\fontsize{10.5pt}{13pt}\selectfont
}

\newpage

% ==================================================
\section{Introduction}
% ==================================================

The relationship between natural resource extraction and armed conflict is an important issue in development economics and international policy. To sever this relationship, Section 1502 of the Dodd--Frank Act was enacted in the United States in 2010, requiring U.S.-listed companies to disclose information concerning their use of tin, tantalum, tungsten, and gold (3TG) originating from the Democratic Republic of the Congo and adjoining countries. Rwanda is one of the adjoining countries covered by the regulation, and its tungsten exports have also been affected by international conflict minerals regulation and due diligence systems. \\
Research on conflict minerals regulation, including Section 1502 of the Dodd--Frank Act, has primarily examined its social and economic impacts. Parker et al. (2016) show that, after the introduction of conflict minerals regulation, infant mortality increased substantially in villages near mines covered by the regulation, and they attribute this in part to reduced household income and the associated decline in spending on and use of health services for infants. Hanai (2021) argues that, although conflict minerals regulation changed the behavior of firms, armed groups, and other actors, the underlying mechanisms linking mineral trade to armed conflict remained in place. Koch and Kinsbergen (2018) further point out that, despite the negative effects of conflict minerals regulation diminishing over time, the negative interpretation of the regulation as a ``de facto embargo'' remained dominant. Stoop et al. (2018) show that, after the introduction of Section 1502 of the Dodd--Frank Act, the incidence of fighting, looting, and violence against civilians increased, particularly in areas with many gold mines, suggesting that conflict minerals regulation may have unintentionally worsened local violence even over the longer term. \\
While these studies examine the social and regional economic impacts of conflict minerals regulation, other studies analyze the effects of regulation and certification systems on mineral trade itself. Using international trade data from 2006 to 2017, Sch\"{u}tte (2019) shows that imports of tantalum and tin from the Great Lakes region temporarily declined after the introduction of the Dodd--Frank Act and due diligence systems but subsequently recovered. The study also finds diversification of importing countries for tantalum and improved consistency in trade statistics alongside the implementation of due diligence. However, the study does not cover tungsten and does not analyze Rwanda-specific trade patterns or changes in the price elasticity of export demand.\\
Regarding the relationship between certification systems and international trade, Borsky and Leiter (2022) analyze the impact of the Kimberley Process Certification Scheme (KPCS) on international trade in rough diamonds using bilateral trade data from 1996 to 2015 and a structural gravity model. Their results show that when both exporters and importers participate in the KPCS, improved information transparency and common standards increase trade values and the probability of trade, whereas when only the exporter participates, compliance costs may reduce export competitiveness. Although this study demonstrates that the effects of certification systems on international trade are not unidirectional, it focuses on diamonds and does not examine changes in export demand price responses or price elasticities of Rwandan tungsten, which is covered by the Dodd--Frank Act, before and after the introduction of the regulation.\\
With respect to methods for estimating trade elasticities, Nakano and Nishimura (2025) present a framework that identifies trade elasticities using exchange-rate fluctuations when prices and quantities cannot be sufficiently observed. For Rwandan tungsten as well, it is difficult to stably observe export quantities and export prices throughout the full sample period. Therefore, following this approach, this study does not directly use observed unit values as prices but instead identifies export demand elasticity using exchange-rate fluctuations.\\
Furthermore, Nagamori and Nishimura (2026) estimate the price elasticity of export demand for tin from the DRC before and after the introduction of the Dodd--Frank Act. They find a high degree of price responsiveness before the regulation, whereas the price response disappears after implementation of the regulation and remains absent even after the SEC's change in enforcement policy in 2017. They also report that price responsiveness reappears after the 2019 Huawei shock, suggesting a possible recovery in market competitiveness.\\
As described above, previous studies have examined a wide range of effects of conflict minerals regulation on people's livelihoods, the behavior of firms and armed groups, regional economies, and international trade. Research on certification systems and international trade has also suggested that such systems may promote trade through improved information transparency while simultaneously reducing export competitiveness through compliance costs. In addition, changes in the price elasticity of export demand before and after the introduction of the Dodd--Frank Act have been analyzed for tin from the DRC.\\
However, it remains unclear how the price responsiveness of export demand for Rwandan tungsten changed over time under the introduction of the Dodd--Frank Act and subsequent changes in the institutional environment. In particular, it has not been sufficiently examined whether changes following the introduction of the regulation persisted or whether they changed again in later periods.\\
Accordingly, this study uses monthly mirror trade data on Rwandan tungsten ores and concentrates from January 2009 to December 2023 and estimates the price elasticity of export demand across four periods: before full-scale implementation of the regulation, during the implementation period of the Dodd--Frank Act regulation, after the April 2017 change in enforcement policy by the U.S. Securities and Exchange Commission (SEC), and during the shift in export destinations from 2019 onward. The purpose of this study is to clarify how the demand elasticity of Rwandan tungsten exports evolved over time, considering not only the periods before and after implementation of the regulation under the Dodd--Frank Act but also subsequent changes in the institutional environment. This enables an examination of whether the changes in price responsiveness observed for DRC tin also appear for Rwandan tungsten, which is subject to the same regulation. Among the 3TG minerals, tungsten production in the covered region is almost entirely concentrated in Rwanda; therefore, this study focuses on Rwandan tungsten.\\

% ==================================================
\section{Data and Methodology}
% ==================================================

\subsection{Sample Period}

The sample period runs from January 2009 to December 2023 and is divided into the following four periods based on changes in the institutional environment and the composition of export destinations.

\begin{itemize}
    \item Period 1: January 2009--December 2012
    \item Period 2: January 2013--April 2017
    \item Period 3: May 2017--May 2019
    \item Period 4: June 2019--December 2023
\end{itemize}

Period 1 is the period before full-scale implementation of conflict minerals regulation under the Dodd--Frank Act. UN Comtrade trade data show that Rwandan tungsten exports during this period were centered on China, and this pattern continued broadly through the end of 2012. The SEC adopted the conflict minerals rule in August 2012, and firms were required to comply with the rule beginning with calendar year 2013. Therefore, this study treats the period through December 2012 as the pre-implementation baseline period.

Period 2, from January 2013 to April 2017, is defined as the implementation period of conflict minerals regulation under the Dodd--Frank Act. Because the SEC announced a change in its enforcement policy in April 2017, that month is included in Period 2, while Period 3 begins in the following month, May 2017.

Period 3, from May 2017 to May 2019, is defined as the period following the SEC's change in enforcement policy. In April 2017, in light of a court ruling, the SEC changed its enforcement treatment of certain parts of the conflict minerals rule (U.S. Securities and Exchange Commission, 2017). Although this did not repeal Section 1502 of the Dodd--Frank Act itself, it can be regarded as a point at which the regulatory environment faced by firms changed; therefore, this period is analyzed separately from Period 2.

Period 4 is defined as June 2019 through December 2023, based on the observed change in the composition of destinations for Rwandan tungsten exports around 2019 in UN Comtrade trade data. To examine whether the price responsiveness of export demand differs before and after this change in export-destination composition, this study separates the period through May 2019 from the period beginning in June 2019.

\begin{figure}[H] 
\centering 
\includegraphics[width=0.9\textwidth]{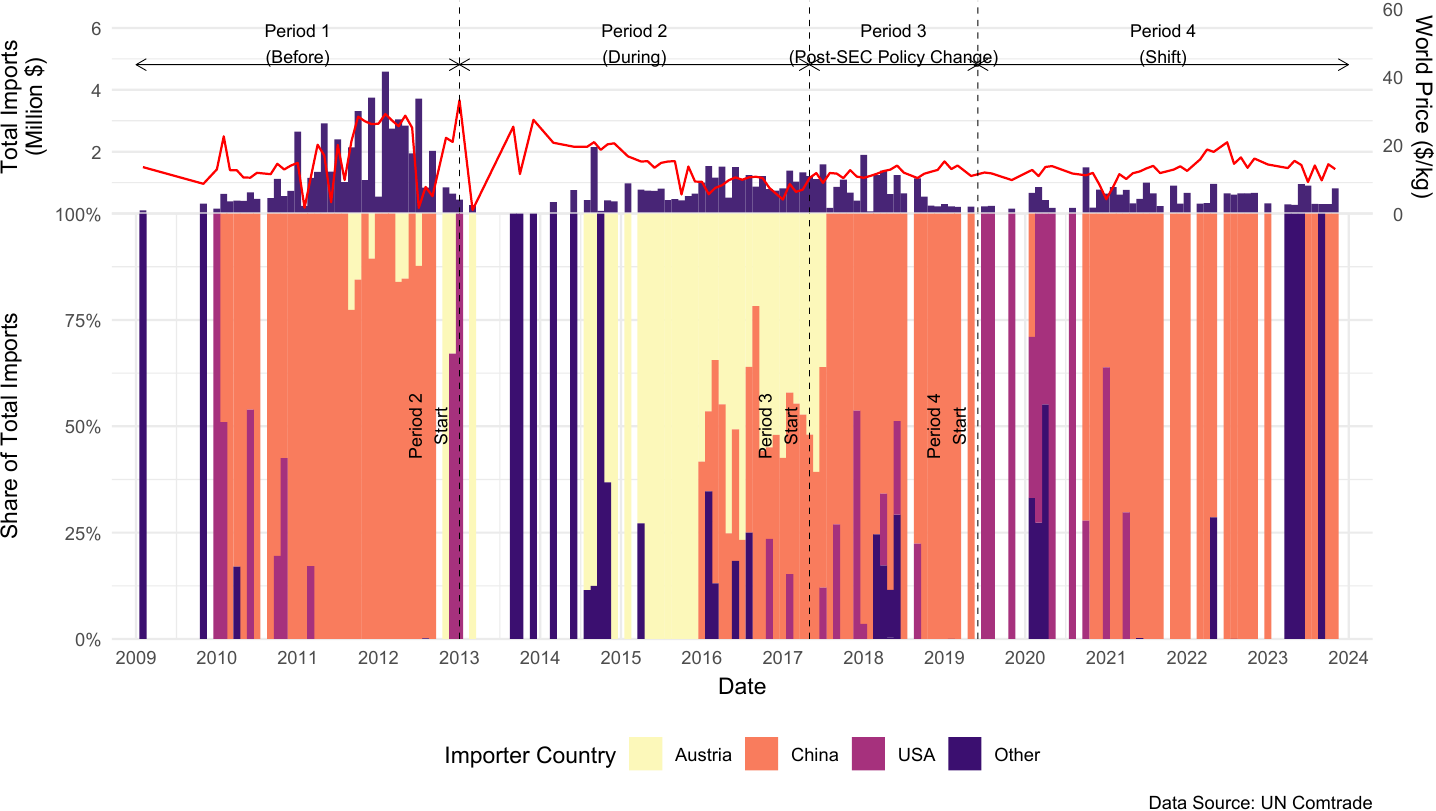}
\caption{Rwandan Tungsten Export Values, World Price, and Composition of Export Destinations} \label{fig:tungsten} \end{figure}

Figure \ref{fig:tungsten} shows the evolution of monthly export values of Rwandan tungsten (based on importer reports), the world tungsten price, and the composition of export destinations. The upper panel shows monthly total export values and the world price, while the lower panel shows the shares of import values accounted for by Austria, China, the United States, and other countries. The four sample periods defined in this study based on trade data are also indicated. The composition of export destinations is not constant throughout the sample period, and the main destinations can be seen to change over time.

\subsection{Data and World Average Price}

This study uses importer-reported trade data for Rwandan tungsten obtained from UN Comtrade, as well as global import data used to construct the world average price of tungsten. The product category is HS code 261100 (tungsten ores and concentrates) in both cases. For exchange rates, monthly exchange-rate data constructed from FXTOP are used. Based on these data, the world average price is constructed and merged, together with exchange-rate data, into the panel dataset used for the analysis.

For the world tungsten price, monthly import data for HS code 261100 (tungsten ores and concentrates) obtained from UN Comtrade are used. The world average price $P_t^W$ in month $t$ is calculated by dividing the sum of import values from the world reported by each importing country by the sum of corresponding import quantities, as follows.

\begin{equation}
P_t^{W}
=
\frac{\sum_i V_{it}}
     {\sum_i Q_{it}}
\end{equation}

Here, $V_{it}$ denotes the import value of importing country $i$ in month $t$, and $Q_{it}$ denotes the corresponding import quantity. This indicator is equivalent to a monthly world average price in which each country's import unit value is weighted by import quantity. The world average price is not a variable used to directly calculate the export demand elasticity; rather, it is included in the estimating equation to capture price fluctuations common to the global tungsten market.

In the monthly UN Comtrade trade data used in this study, export values are observed relatively widely, whereas export quantities are missing for some observations, making it difficult to secure quantity data that are consistently available over the full sample period. Therefore, if unit values obtained by dividing export values by quantities were used as export prices, the sample would be restricted to the subset of transactions for which quantities are observed, making it difficult to compare price responses across periods within a common framework. Given this data constraint, this study applies the approach of Nakano and Nishimura (2025), which identifies trade elasticities using exchange-rate fluctuations even when physical trade quantities and prices are not sufficiently observed.

\subsection{Estimation Model}

This study assumes that Rwanda is a small country in the world tungsten market and is therefore a price taker that accepts the world market price as given. Export demand for importing country $j$ at time $t$ is expressed as follows.

\begin{equation}
\ln Q_{jt}
=
\alpha_j
+
\eta \ln P^{imp}_{jt}
+
u_{jt}
\end{equation}

Here, $Q_{jt}$ denotes export quantity, $P^{imp}_{jt}$ denotes the price in the importing country, and $\eta$ is the price elasticity of export demand.

Because export quantities and export prices cannot be stably observed over the entire sample period, this study uses export value $X_{jt}=P^{imp}_{jt}Q_{jt}$. The price in the importing country is expressed as

\begin{equation}
\ln P^{imp}_{jt}
=
\ln E_{jt}
+
\ln P^{exp}_{jt}
+
\tau_j
\end{equation}

where $E_{jt}$ is the exchange rate expressed as importing-country currency per Rwandan franc, $P^{exp}_{jt}$ is the export price from Rwanda to importing country $j$, and $\tau_j$ represents importer-specific and time-invariant factors such as transportation costs, tariffs, and trade routes.

The export price is affected by the world market price but may also respond to changes in bilateral export values. Accordingly, the export-price response is expressed as follows.

\begin{equation}
\ln P^{exp}_{jt}
=
\delta_j
+
\omega \ln X_{jt}
+
\lambda \ln P^{W}_{t}
+
v_{jt},
\end{equation}

where $P^{W}_{t}$ is the world tungsten price, $\delta_j$ is a time-invariant price differential specific to the importing country or trade route, and $v_{jt}$ is a supply-side shock.

Export prices may be determined through the interaction of demand and supply. If they are correlated with demand shocks, directly using export prices as an explanatory variable may generate endogeneity due to simultaneity. Therefore, this study combines the demand equation and the price-response equation to eliminate the export price and uses the following reduced-form equation.

\begin{equation}
\ln X_{jt}
=
\mu_j
+
\beta \ln E_{jt}
+
\gamma \ln P^{W}_{t}
+
\varepsilon_{jt},
\end{equation}

where $\mu_j$ is an importer fixed effect. The relationship between the exchange-rate coefficient $\beta$ and the price elasticity of export demand $\eta$ is given by

\begin{equation}
\beta
=
\frac{1+\eta}
{1-\omega(1+\eta)}
\end{equation}

Therefore, the price elasticity of export demand is generally recovered as

\begin{equation}
\eta
=
\frac{\beta}
{1+\beta\omega}
-1.
\end{equation}

For this reason, this study separately estimates the inverse-supply coefficient $\omega$ using observations for which export quantities are available in order to assess the extent to which export prices respond to bilateral export values. The estimation results show that $\omega$ is not statistically significantly different from zero at the 5\% level and that its point estimate is also small. Based on these results, this study approximates $\omega=0$.

When $\omega=0$, the relationship between the exchange-rate coefficient $\beta$ and the price elasticity of export demand $\eta$ simplifies to

\begin{equation}
\eta = \beta - 1.
\end{equation}

Accordingly, the following analysis uses this relationship to calculate the export demand elasticity for each period. The estimation method and detailed results for $\omega$ are presented in the Appendix.

\section{Estimation Results}

% ==================================================
\begin{table}[htbp]
\centering
\caption{Estimation Results of the Fixed-Effects Model by Period}
\label{tab:fe_results}
\resizebox{\textwidth}{!}{%
\begin{tabular}{lcccc}
\toprule
& Period 1 & Period 2 & Period 3 & Period 4 \\
\midrule

Estimation period
& Jan. 2009--Dec. 2012
& Jan. 2013--Apr. 2017
& May 2017--May 2019
& Jun. 2019--Dec. 2023 \\

\addlinespace

Exchange-rate coefficient $\beta$
& $-19.814^{***}$
& $2.814$
& $-4.277^{***}$
& $1.440$ \\

& $(4.536)$
& $(2.315)$
& $(0.381)$
& $(0.913)$ \\

\addlinespace

World-price coefficient $\gamma$
& $0.307$
& $0.354^{*}$
& $-0.697$
& $-0.036$ \\

& $(0.159)$
& $(0.156)$
& $(1.062)$
& $(0.182)$ \\

\midrule

Export demand elasticity $\eta$
& $-20.814^{***}$
& $1.814$
& $-5.277^{***}$
& $0.440$ \\

95\% confidence interval
& $[-29.704,\,-11.924]$
& $[-2.723,\,6.351]$
& $[-6.024,\,-4.531]$
& $[-1.350,\,2.229]$ \\

\addlinespace

Observations
& 52 & 64 & 37 & 49 \\

Number of importing countries
& 4 & 7 & 4 & 5 \\

\bottomrule
\end{tabular}%
}

\begin{flushleft}
\footnotesize
Note: Standard errors clustered at the importing-country level are shown in parentheses. Importer fixed effects are included in all estimations. Export demand elasticity is calculated as $\eta \approx \beta - 1$. $^{***}p<0.01$, $^{**}p<0.05$, $^{*}p<0.10$.
\end{flushleft}

\end{table}

\subsection{Period-by-Period Estimation Results}

Period 1: Before implementation of the regulation \\
Period 1 covers January 2009 to December 2012, before implementation of the regulation. The estimated export demand elasticity in this period is $-20.814$ and is statistically significant at the 1\% level. This indicates that, before implementation of the regulation, export demand for Rwandan tungsten responded strongly to price changes.

Period 2: Dodd--Frank period\\
Period 2 covers January 2013 to April 2017 and corresponds to the Dodd--Frank period. The estimated export demand elasticity in this period is 1.814 and is not statistically significant. Therefore, no clear export-demand response to price changes is observed in this period.

Period 3: After the SEC's change in enforcement policy \\
Period 3 covers May 2017 to May 2019, following the SEC's change in enforcement policy. The estimated export demand elasticity in this period is $-5.277$ and is statistically significant at the 1\% level. Therefore, a negative export-demand response to price changes is observed in this period.

Period 4: Shift in export destinations\\
Period 4 covers June 2019 to December 2023 and corresponds to the period of a shift in export destinations. The estimated export demand elasticity in this period is 0.440 and is not statistically significant. Therefore, no clear export-demand response to price changes is observed in this period.

\subsection{Coefficient Differences across Periods}

Statistical significance of the coefficients within individual periods alone does not determine whether the coefficients themselves differ statistically across periods. Therefore, this study conducts coefficient-difference tests to examine changes in price responsiveness before and after implementation of the regulation and in subsequent periods. In particular, the data show an explicit change in tungsten export destinations between Periods 3 and 4. However, because the estimation results show a negative and statistically significant price response in Period 3 but no statistically significant price response in Period 4, the difference between the coefficients in Periods 3 and 4 is also tested to determine whether the coefficient itself changed statistically at the June 2019 boundary.

The coefficient-difference tests show that the difference between Period 1 and Period 2 is statistically significant at the 1\% level ($p=0.0021$). The differences between Period 1 and Period 3 ($p=0.0097$) and between Period 1 and Period 4 ($p<0.001$) are also statistically significant at the 1\% level. Thus, the price response in Period 1 is statistically different from those in each of the subsequent periods.

The coefficient difference between Period 3 and Period 4 is also statistically significant at the 1\% level ($p<0.001$). Thus, the price responses in Periods 3 and 4 are statistically different.

% ==================================================
\section{Discussion}
% ==================================================

In Period 1, the export demand elasticity is large and negative, indicating a clear export-demand response to price changes. This period precedes implementation of conflict minerals regulation under the Dodd--Frank Act, and it has been reported that multiple firms were involved in tungsten production in Rwanda during Period 1 (Yager, 2014). This suggests that trade routes may have been less fixed than in the subsequent regulation period, and such a market environment may have made transaction adjustments in response to prices more likely. However, direct evidence of competition among buyers or switching of suppliers is limited, and the absolute magnitude of the estimate is also large; therefore, the size of the estimate itself should be interpreted with caution.

In Period 2, the export demand elasticity is not statistically significant, and no clear export-demand response to price changes is observed. The coefficient difference from Period 1 is also statistically significant, indicating a price response different from that before implementation of the regulation. This period corresponds to implementation of conflict minerals regulation under the Dodd--Frank Act and overlaps with a time when some international buyers avoided sourcing minerals from the Great Lakes region while due diligence practices, including traceability and certification systems, expanded (Sch\"{u}tte, 2019). These institutional factors may have increased the importance of non-price conditions in trade, which may explain why the price response observed in Period 1 is no longer clearly observed in Period 2. However, direct evidence of buyer exit or the fixation of trading networks for Rwandan tungsten is limited, and the sample size is not large; therefore, it cannot be concluded that the Dodd--Frank Act directly caused the change in price responsiveness.

In Period 3, the export demand elasticity is negative and statistically significant, and the export-demand response to price changes that was not observed in Period 2 reappears. This period follows the SEC's April 2017 change in enforcement policy for the conflict minerals rule and also exhibits changes in the composition of export destinations, including an increased share of exports to China. These changes in the institutional and market environment are consistent with the reemergence of transaction adjustments in response to price conditions. However, it cannot be concluded that the SEC's enforcement-policy change or the increase in trade with China directly generated the price response. Moreover, the coefficient difference between Period 1 and Period 3 is statistically significant, meaning that the reappearance of price responsiveness in Period 3 does not imply a return to the pre-regulation state.

In Period 4, the export demand elasticity is not statistically significant, and the negative price response observed in Period 3 is no longer observed. The coefficient difference between Period 3 and Period 4 is statistically significant, confirming that price responsiveness differs between the two periods. Figure 1 shows that the composition of export destinations also changed around 2019. In Period 4, exports to the United States are observed, while the share of exports to China is not constant throughout the period. Thus, the transition from Period 3 to Period 4 was not only a change in price responsiveness but also a period in which the composition of Rwandan tungsten's trading partner countries changed. Changes in the composition of export destinations may have altered demand conditions and trading relationships in each importing country, and such changes in market composition may be related to the difference in price responsiveness observed in Period 4.

% ==================================================
\section{Conclusion}
% ==================================================

\subsection{Main Findings and Contributions}

This study estimates the price elasticity of export demand for Rwandan tungsten exports from January 2009 to December 2023 across four periods.

The estimation results show that in Period 1 the export demand elasticity is negative and statistically significant, indicating a clear export-demand response to price changes. In contrast, in Period 2 the export demand elasticity is not statistically significant, and no clear price response is observed. In Period 3, the export demand elasticity becomes negative and statistically significant again, and the price response not observed in Period 2 reappears. In Period 4, however, the export demand elasticity is not statistically significant, and the price response observed in Period 3 does not persist.

The coefficient-difference tests also identify statistically significant differences between Period 1 and Period 2, Period 1 and Period 3, Period 1 and Period 4, and Period 3 and Period 4. Taken together, these results show that the price responsiveness of Rwandan tungsten exports did not change in only one direction after the introduction of regulation under the Dodd--Frank Act, but instead varied over time, including in subsequent periods.

The first contribution of this study is that it examines the evolution of the price elasticity of export demand for Rwandan tungsten not only before and after implementation of regulation under the Dodd--Frank Act but also over the subsequent periods. The estimation results show that the negative price response observed before the regulation is no longer observed during the Dodd--Frank period, reappears after 2017, and is again no longer observed after 2019. In addition, the coefficient differences between the pre-regulation Period 1 and each subsequent period are all statistically significant, indicating that post-regulation price responsiveness evolved in a state different from that before regulation. The significance of this study therefore lies in examining the effects of conflict minerals regulation not only in terms of changes immediately following its introduction but also from the perspective of how market responses evolved thereafter.

The second contribution is the empirical estimation of the price elasticity of export demand for Rwandan tungsten, which has not previously been analyzed in sufficient depth. Although previous studies have examined trade in tin and tantalum from the Great Lakes region, research examining the export-demand response of Rwandan tungsten to price changes is limited. This study provides new empirical evidence for tungsten, a mineral subject to conflict minerals regulation.

The third contribution is to examine whether the changes in price responsiveness documented for DRC tin also appear for Rwandan tungsten, which is subject to the same Dodd--Frank Act. The estimation results show changes in price responsiveness before and after implementation of the regulation, but the subsequent trajectory is not necessarily identical to that reported for DRC tin. This result suggests that market responses under conflict minerals regulation may differ by mineral and exporting country and that caution is required when generalizing findings from a single mineral directly to other 3TG minerals.

\subsection{Limitations and Future Research}

First, this study is based on period-specific fixed-effects estimation and coefficient-difference tests and does not directly identify the causal effect of the Dodd--Frank Act. In addition, the boundary of Period 4 is based on an observed change in the composition of export destinations rather than an exogenous institutional change. Therefore, it cannot be ruled out that changes in the market environment other than regulation affected the changes in price responsiveness observed across periods.

Second, although a statistically significant coefficient difference between Period 3 and Period 4 is identified and a change in the composition of export destinations is also observed during the same period, the relationship between these two changes and the specific factors that produced the change in price responsiveness remain unclear.

Third, the numbers of observations and importing countries in each period are limited, and the estimation results may be influenced by trends in particular importing countries or transactions. Therefore, the magnitude of the elasticity in each period should be interpreted with caution.

For these reasons, the results of this study should not be interpreted as directly demonstrating the causal effect of the Dodd--Frank Act, but rather as showing the evolution of price responsiveness in Rwandan tungsten exports before and after introduction of the regulation and over subsequent periods. Future research should examine transaction structures by export destination and changes in the institutional and market environment in greater detail in order to clarify the factors underlying the changes in price responsiveness observed across periods.

\appendix
\section{Estimation of the Inverse-Supply Coefficient $\omega$}

To assess the extent to which export prices respond to bilateral export values, the inverse-supply coefficient $\omega$ is estimated using observations for which export quantities are available.

The export unit value $UV_{jt}$ for importing country $j$ at time $t$ is defined as

\begin{equation}
UV_{jt}
=
\frac{X_{jt}}{Q_{jt}}
\end{equation}

where $X_{jt}$ is export value and $Q_{jt}$ is export quantity.

The following fixed-effects model is then estimated.

\begin{equation}
\ln UV_{jt}
=
a_j
+
\omega \ln X_{jt}
+
\lambda \ln P^{W}_{t}
+
e_{jt},
\end{equation}

where $a_j$ is an importer fixed effect and $P^{W}_{t}$ is the world tungsten price. Standard errors are clustered at the importing-country level.

The estimate of $\omega$ is 0.04384, with a standard error of 0.02660 and a 95\% confidence interval of $[-0.00829,\,0.09597]$. Thus, $\omega$ is not statistically significantly different from zero at the 5\% level. In addition, because the point estimate is small, the response of export prices to changes in bilateral export values is limited.

Based on these results, this study approximates $\omega=0$.

\section{Coefficient-Difference Tests across Periods}

Statistical significance of the coefficients within individual periods alone does not determine whether the coefficients themselves differ statistically across periods. Therefore, differences in exchange-rate coefficients across periods are tested.

For comparisons between period $r$ and period $s$, a fixed-effects model is estimated using data combining observations from both periods, while allowing the exchange-rate coefficient and the world-price coefficient to differ by period. In addition, fixed effects for importer-by-period combinations are included to allow importer-specific intercepts in each period. Standard errors are clustered at the importing-country level.

The null hypothesis is

\begin{equation}
H_0:\beta_r=\beta_s.
\end{equation}

Defining the coefficient difference as

\begin{equation}
\Delta\beta=\beta_s-\beta_r,
\end{equation}

its standard error is calculated as

\begin{equation}
SE(\Delta\beta)
=
\sqrt{
\mathrm{Var}(\hat{\beta}_s)
+
\mathrm{Var}(\hat{\beta}_r)
-
2\mathrm{Cov}(\hat{\beta}_s,\hat{\beta}_r)
}.
\end{equation}

Using this, the test statistic is calculated as

\begin{equation}
t=
\frac{\hat{\beta}_s-\hat{\beta}_r}
{SE(\Delta\beta)}.
\end{equation}

Because $\eta=\beta-1$ under $\omega \simeq 0$,

\begin{equation}
\eta_s-\eta_r
=
\beta_s-\beta_r,
\end{equation}

so the test of the difference in exchange-rate coefficients is equivalent to a test of the difference in export demand elasticities.

\begin{table}[htbp]
\centering
\caption{Coefficient-Difference Tests across Periods}
\label{tab:coefficient_difference}
\begin{tabular}{lrrrr}
\hline
Comparison & Coefficient difference & Standard error & $t$-value & $p$-value \\
\hline
Period 2 -- Period 1 & 22.628 & 5.665 & 3.994 & 0.0021 \\
Period 3 -- Period 1 & 15.536 & 4.162 & 3.733 & 0.0097 \\
Period 4 -- Period 1 & 21.254 & 4.063 & 5.230 & 0.0008 \\
Period 4 -- Period 3 & 5.717  & 0.960 & 5.955 & 0.0006 \\
\hline
\end{tabular}

\begin{flushleft}
\footnotesize
Note: The coefficient difference is calculated by subtracting the exchange-rate coefficient of the earlier period from that of the later period. For each comparison, a fixed-effects model allowing period-specific exchange-rate and world-price coefficients is estimated, including importer $\times$ period fixed effects. Standard errors are clustered at the importing-country level. Because $\eta=\beta-1$, the difference in exchange-rate coefficients is equal to the difference in export demand elasticities.
\end{flushleft}
\end{table}

\section*{Data Sources}
United Nations (UN) (n.d.).
UN Comtrade Database.
\url{https://comtradeplus.un.org/}.
Accessed Jun 10, 2026.

FXTOP (n.d.).
Historical Exchange Rates.
\url{https://fxtop.com/}.
Accessed Jun 10, 2026.

% ==================================================
\section*{References}
\addcontentsline{toc}{section}{References}
% ==================================================

Borsky, S. and Leiter, A. M. (2022).
International trade in rough diamonds and the Kimberley Process Certification Scheme.
\textit{World Development}, 152, 105786.
\url{https://doi.org/10.1016/j.worlddev.2021.105786}.

Hanai, K. (2021).
Conflict minerals regulation and mechanism changes in the DR Congo.
\textit{Resources Policy}, 74, 102394.
\url{https://doi.org/10.1016/j.resourpol.2021.102394}.

Koch, D.-J. and Kinsbergen, S. (2018).
Exaggerating unintended effects? Competing narratives on the impact of conflict minerals regulation.
\textit{Resources Policy}, 57, 255--263.
\url{https://doi.org/10.1016/j.resourpol.2018.03.011}.

Nagamori, H. and Nishimura, K. (2026).
The impact of Dodd--Frank and the Huawei shock on DRC tin exports.
\textit{Resources Policy}, 117, 105938.
\url{https://doi.org/10.1016/j.resourpol.2026.105938}.

Nakano, S. and Nishimura, K. (2025).
How do we measure trade elasticity for services?
\textit{Empirical Economics}, 68(3), 1477--1494.
\url{https://doi.org/10.1007/s00181-024-02675-z}.

Parker, D. P., Foltz, J. D., and Elsea, D. (2016).
Unintended consequences of sanctions for human rights:
Conflict minerals and infant mortality.
\textit{The Journal of Law and Economics}, 59(4), 731--774.
\url{https://doi.org/10.1086/691793}.

Sch\"{u}tte, P. (2019).
International mineral trade on the background of due diligence regulation:
A case study of tantalum and tin supply chains from East and Central Africa.
\textit{Resources Policy}, 62, 674--689.
\url{https://doi.org/10.1016/j.resourpol.2018.11.017}.

Stoop, N., Verpoorten, M., and van der Windt, P. (2018).
More legislation, more violence?
The impact of Dodd--Frank in the DRC.
\textit{PLOS ONE}, 13(8), e0201783.
\url{https://doi.org/10.1371/journal.pone.0201783}.

U.S. Securities and Exchange Commission (SEC) (2017).
Updated Statement on the Effect of the Court of Appeals Decision on the Conflict Minerals Rule.
Division of Corporation Finance, April 7, 2017.
\url{https://www.sec.gov/newsroom/speeches-statements/corpfin-updated-statement-court-decision-conflict-minerals-rule}

Yager, T. R. (2014).
The Mineral Industry of Rwanda in 2012.
\textit{U.S. Geological Survey Minerals Yearbook 2012}.
U.S. Geological Survey.

\end{document}